\documentclass[12pt]{article}
\usepackage{amssymb,amsmath}
\begin{document}

\begin{flushright}
{\tt hep-th/0501016}
\end{flushright}

\vspace{5mm}

\begin{center}
{{{\Large \bf Exact Rolling Tachyon\\[2mm]
in Noncommutative Field Theory}}\\[14mm]
{Yoonbai Kim~~and~~O-Kab Kwon}\\[2.5mm]
{\it BK21 Physics Research Division and Institute of
Basic Science,\\
Sungkyunkwan University, Suwon 440-746, Korea}\\
{\tt yoonbai@skku.edu~~okab@skku.edu} }
\end{center}
\vspace{10mm}

\begin{abstract}
We study the exact rolling tachyon solutions in DBI type
noncommutative field theory with a constant open string metric and
noncommutative parameter on an unstable D$p$-brane.
Functional shapes of the obtained solutions span all possible
homogeneous rolling tachyon configurations;
that is, they are hyperbolic-cosine, hyperbolic-sine, and
exponential under $1/\cosh$ runaway NC tachyon potential. Even if
general DBI type NC electric field is turned on, only a constant
electric field satisfies the equations of motion, and again, exact
rolling tachyon solutions are obtained.
\end{abstract}

\newpage

\setcounter{equation}{0}
\section{Introduction}

Rolling tachyon was constructed as classical time dependent solutions in open
string theory, describing real-time dynamics of tachyonic degree on
unstable D-branes~\cite{Sen:2002nu,Sen:2002an,Sen:2004nf}.
When NSNS two-form field and the gauge field
exist on the D-brane, two descriptions of effective field theory (EFT)
can be applied:
One is the usual EFT coupled to Dirac-Born-Infeld (DBI)
electromagnetism and the other is noncommutative (NC)
field theory (NCFT)~\cite{Seiberg:1999vs}.

When a constant electric component of the NSNS two-form field or an equivalently
constant electric field is turned on,
various approaches in terms of boundary conformal field theory
(BCFT)~\cite{Mukhopadhyay:2002en},
EFT~\cite{Mukhopadhyay:2002en,Gibbons:2002tv}, and
NCFT~\cite{Mukhopadhyay:2002en}
are employed to obtain rolling tachyon solutions which are compared
at the level of energy-momentum tensor mostly for large elapsed time.
At this new vacuum, tachyon matter~\cite{Sen:2002nu}
does not contribute  to
pressure but fundamental strings (F1s)  in the fluid
state~\cite{Gibbons:2000hf}
do~\cite{Mukhopadhyay:2002en,Gibbons:2002tv}. For any runaway NC
tachyon potentials, three species of rolling tachyon solutions can be found,
which are identified as two full S-brane solutions and one half S-brane
solution~\cite{Gutperle:2002ai,Maloney:2003ck,Hashimoto:2003qx}.

When both electric and magnetic components of the NSNS two-form field are
taken into account, the same three types of rolling tachyon solutions are
obtained in BCFT~\cite{Rey:2003xs} and
EFT~\cite{Kim:2004zq} irrespective of various constant electromagnetic field
configurations and specific shapes of tachyon potentials.
In the EFT of DBI action with $1/\cosh$ type tachyon potential,
homogeneous rolling tachyon solutions are given as exact
solutions despite of nonlocality and
many field strength components for arbitrary
D$p$-brane~\cite{Kim:2003he,Kim:2003ma}.

NC electromagnetic field and NC tachyon with
its condensation have been an attractive subject since initial
investigations on NCFT~\cite{Dasgupta:2000ft}. Studies on exact
tachyon kinks have been performed recently~\cite{Banerjee:2004cw,Kim:2004xn},
but exact homogeneous
rolling tachyon solutions have yet to be reported in the context of NCFT.
In this paper we consider an NC DBI action and $1/\cosh$ type NC tachyon
potential with constant open string metric and NC parameter,
and investigate exact homogeneous NC rolling tachyon solutions.
We attempt to find such solutions for cases with or without
coupling of the NC electromagnetic
field, and they had one-to-one correspondence with those in EFT.
For their energy-momentum tensor, two full S-brane solutions are in agreement
qualitatively to but half S-brane solution coincides exactly with
that of BCFT.

The rest of this paper is organized as follows.
In section 2, the most general homogeneous solution of NC rolling tachyon with
open string metric and NC parameter is derived from a constant NSNS two-form field.
In section 3, DBI NC electromagnetic field is added and we also find
homogeneous NC rolling tachyon solution for the case with
the NC electric field.
In section 4, results and conclusions are presented.

\setcounter{equation}{0}
\section{NC Rolling Tachyon}

In this section, we study rolling tachyon in NCFT, describing dynamics
of an unstable
D$p$-brane coupled to both constant
DBI type electromagnetic field and constant antisymmetric NSNS two-form
field. All possible homogeneous NC rolling tachyons
are found as exact solutions.
In the context of EFT, rolling tachyons with DBI type
electromagnetism~\cite{Mukhopadhyay:2002en,Gibbons:2002tv}
is described by the action
\begin{equation}\label{fa}
S= -{\cal T}_p \int d^{p}x\; V(T) \sqrt{-\det (\eta_{\mu\nu} +
F_{\mu\nu}+\partial_\mu T\partial_\nu T)}\, ,
\end{equation}
and, for specific form of tachyon
potential~\cite{Buchel:2002tj,Kim:2003he,Kutasov:2003er},
\begin{equation}\label{V3}
V(T)=\frac{1}{\cosh \left(\frac{T}{R}\right)}.
\end{equation}
All possible homogeneous rolling tachyons are obtained as exact
solutions~\cite{Kim:2003he,Kim:2004zq}. Specifically, for an arbitrary $p$,
every component of the field strength tensor $F_{\mu\nu}$ satisfying the gauge
equation and Bianchi identity should be constant~\cite{Kim:2003ma},
and then the tachyon profiles, $T(t)$, are obtained by solving
a first-order equation given by the definition of the constant Hamiltonian density
${\cal H}$
\begin{equation}\label{eqn}
{\cal H}=\frac{\alpha_{p0}{\cal T}_{p}V(T)}{
\sqrt{\beta_{p}-\alpha_{p0}\dot{T}^{2}}},
\end{equation}
where $\beta_{p}=-\det (\eta_{\mu\nu}+F_{\mu\nu})$ and $\alpha_{p0}>0$ is
a 00-component of the cofactor $C^{\mu\nu}$ of matrix
$\eta_{\mu\nu} + F_{\mu\nu}+ \partial_\mu T\partial_\nu T$.
For $\beta_{p}-\alpha_{p0}\dot{T}^{2}$, $\beta_{p}>0$ in order to obtain
 nontrivial rolling tachyon solutions. This condition leads to critical
 value of the electromagnetic field.
Homogeneous rolling tachyon solutions exist for positive $\beta_{p}$
and, by value of the Hamiltonian density ${\cal H}$, solutions of
Eq.~(\ref{eqn}) are classified into three forms,
\begin{equation}\label{rss3}
\sinh\left(\frac{T(t)}{R}\right)=
\left\{
\begin{array}{cl}
\sqrt{u^{2}-1} \cosh\left(t/\zeta\right) &
\mbox{for}~{\cal H}<\sqrt{\frac{\alpha_{p0}}{\beta_{p}}}{\cal T}_{p} \\
\exp \left(t/\zeta\right)
&\mbox{for}~{\cal H}=\sqrt{\frac{\alpha_{p0}}{\beta_{p}}}{\cal T}_{p}\\
\sqrt{1-u^2} \sinh\left(t/\zeta\right)
&\mbox{for}~{\cal H}>\sqrt{\frac{\alpha_{p0}}{\beta_{p}}}{\cal T}_{p}
\end{array}
\right. ,
\end{equation}
where $u=\alpha_{p0}{\cal T}_{p}/(\beta_{p}{\cal H})$ and
$\zeta=\sqrt{\alpha_{p0}/\beta_{p}}\, R$.

Since the electromagnetic field $F_{\mu\nu}$ of interest is constant
on the D-brane and gauge invariant configuration on this D-brane
is given by the sum of the
electromagnetic field and antisymmetric NSNS two-form field,
$B_{\mu\nu}+F_{\mu\nu}$,
one can regard the constant electromagnetic field as a part of constant
NSNS two-form field on the D-brane.
When we have an open string theory with flat closed string metric
$\eta_{\mu\nu}$ and constant background two-form field
$B_{\mu\nu}=F_{\mu\nu}$,
a BCFT calculation of the propagator on a disc, which corresponds to a
point splitting regularization of string theory, provides an open
string metric $G^{\mu\nu}$ and NC parameter $\theta^{\mu\nu}$ in
terms of the closed string variables as
\begin{eqnarray}
G_{\mu\nu}&=&g_{\mu\nu}-(Bg^{-1}B)_{\mu\nu},
\label{omet}\\
\theta^{\mu\nu}&=&-\left(\frac{1}{g+B}B\frac{1}{g-B}\right)^{\mu\nu}.
\label{ncpa}
\end{eqnarray}
According to the result of Ref.~\cite{Seiberg:1999vs}, the DBI action
(\ref{fa}) without the tachyon is equivalent to NC DBI action $S_{\Theta}$:
\begin{equation}\label{SW}
S(g_{\mu\nu},B_{\mu\nu};T=F_{\mu\nu}=0)=
\sqrt{\frac{-\det (g_{\mu\nu}+B_{\mu\nu})}{-\det G_{\mu\nu}}}
\, S_{\Theta}(G_{\mu\nu},\theta^{\mu\nu};{\hat T}={\hat F}_{\mu\nu}=0),
\end{equation}
where every product in $S_{\Theta}$ is replaced by star product $(\ast)$
\begin{equation}
f(x)\ast g(x)\equiv e^{\frac{i}{2}\theta^{\mu\nu}
\frac{\partial}{\partial\xi^{\mu}}\frac{\partial}{\partial\zeta^{\nu}}}
f(x+\xi )g(x+\zeta)|_{\xi=\zeta=0}.
\end{equation}
Here, ${\hat T}(x)$ is the NC tachyon and ${\hat F}_{\mu\nu}(x)$ is (usually
slowly varying) the NC electromagnetic field strength tensor
defined by
\begin{equation}\label{ncF}
\hat{F}_{\mu\nu}=\partial_{\mu}\hat{A}_{\nu}-\partial_{\nu}\hat{A}_{\mu}
-i\hat{A}_{\mu}\ast \hat{A}_{\nu} + i\hat{A}_{\nu}\ast
\hat{A}_{\mu}.
\end{equation}

When the NC tachyon
with arbitrary spacetime dependence is taken into account, the NC tachyon
${\hat T}$ is related to the tachyon $T$ by a field redefinition,
and the equivalence between the DBI type EFT action
and the DBI type NC action in Eq.~(\ref{SW}) is not proved, yet.
For the homogeneous rolling tachyon solution depending
only on the time coordinate, every star product in $S_{\Theta}$
is replaced by an ordinary product as
\begin{equation}\label{ncor}
\left. {\hat T}(t)\ast{\hat T}(t)
=e^{\frac{i}{2}\left[2\theta^{tx_{i}}
(\partial_{t}^{\xi}\partial^{\zeta}_{x_{i}}-
\partial_{x_{i}}^{\xi}\partial^{\zeta}_{t})
+\theta^{x_{i}x_{j}}(\partial_{x_{i}}^{\xi}\partial^{\zeta}_{x_{j}}-
\partial_{x_{j}}^{\xi}\partial^{\zeta}_{x_{i}})\right]}{\hat T}(t+\xi)
{\hat T}(t+\zeta)\right|_{\xi=\zeta=0}
={\hat T}^{2}(t),
\end{equation}
and the field redefinition
between ${\hat T}$ and $T$ is identity
by the result of Ref.~\cite{Mukhopadhyay:2002en}
\begin{equation}\label{ide}
{\hat T}(t)=T(t).
\end{equation}
When ${\hat F}_{\mu\nu}=0$, a resultant NC version of the action (\ref{fa}) is
\begin{equation}\label{nct}
{\hat S}=-\frac{{\hat {\cal T}}_p}{2} \int d^{p+1}x\;
\left[\hat{V}(\hat{T}) \ast \sqrt{ -\det {}_{\ast}(G_{\mu\nu}+
\partial_{\mu}\hat{T}\ast\partial_{\nu}\hat{T})}
+(\sqrt{\hspace{-5mm}}\leftrightarrow \hat{V}) \right],
\end{equation}
where ${\hat {\cal T}}_p={\cal T}_{p}/\sqrt{\det(1+g^{-1}B)}\, $.
In fact, the NC action (\ref{nct}) is equivalent to the effective action $S$
up to leading $\theta^{\mu\nu}$ order~\cite{Banerjee:2004cw}.
Because $\hat{V}(\hat{T})\ast \Rightarrow V({\hat T})$ (see
Eq.~(\ref{V3})), $\det {}_{\ast}\Rightarrow\det$, and
$\partial_{\mu}\hat{T}\ast\partial_{\nu}\hat{T}
=\partial_{\mu}{\hat T}\partial_{\nu}{\hat T} =\delta_{\mu
0}\delta_{\nu 0}\dot{{\hat T}}^{2}$ from
Eqs.~(\ref{ncor})--(\ref{ide}), our action (\ref{nct}) with a flat metric
$g_{\mu\nu}\Rightarrow \eta_{\mu\nu}$ is
simplified for a homogeneous solution
\begin{eqnarray}
{\hat S}&=& \sqrt{\frac{-\det (\eta_{\mu\nu}+B_{\mu\nu})}{-\det G_{\mu\nu}}}
\, S_{\Theta}(G_{\mu\nu}, \theta^{\mu\nu};\hat T, \hat F_{\mu\nu}=0)
\label{equ} \\
&=&-\frac{{\cal T}_p}{(-G)^{\frac{1}{4}}}
\int d^{p+1}x\, V(\hat{T})  \sqrt{ -\det (G_{\mu\nu}+
\partial_{\mu}\hat{T} \partial_{\nu}\hat{T})}
\label{nct2} \\
&=& -\frac{{\cal T}_{p}}{({\hat \beta}_{p})^{\frac{1}{4}}}
\int d^{p+1}x\, V({\hat T})
\sqrt{{\hat \beta}_{p}-{\hat \alpha}_{p0}\dot{{\hat T}}^{2}}\, ,
\label{sia2}
\end{eqnarray}
where ${\hat \beta}_{p}=-G=-\det (G_{\mu\nu})\ge 0$,
${\hat \alpha}_{p0}>0$ is $00$-component of the cofactor
${\hat C}^{\mu\nu}$
of matrix $({\hat X})_{\mu\nu}=G_{\mu\nu}+ \partial_\mu \hat T
\partial_\nu \hat T$, and $-\hat X \equiv -\det\hat X_{\mu\nu}={\hat \beta}_{p}
-{\hat \alpha}_{p0}\dot{{\hat T}}^{2}$.

NC energy-momentum tensor is read through variation of the open string metric
\begin{eqnarray}\label{emco}
\hat{T}^{\mu\nu}
&\equiv &  \frac{2}{\sqrt{-G}}
\frac{\delta\hat{S}_\Theta}{\delta G_{\mu\nu}}
\nonumber \\
&=&\frac{ {\cal T}_p  V}{ \sqrt{-G}
\sqrt{-\hat X}}\left(\hat X G^{\mu\nu} - G \partial^{\mu}\hat T
\partial^{\nu}\hat T\right),\label{emte}
\end{eqnarray}
where the tensor indices are lowered and raised by using
the open string metric as $\partial^{\mu}=G^{\mu\nu} \partial_\nu$.
The time-component of the conservation of the energy-momentum tensor
\begin{eqnarray}\label{ncc}
\hat{D}_{\mu} \hat{T}^{\mu\nu}=0
\end{eqnarray}
forces NC Hamiltonian density to be a constant
\begin{equation}\label{nceq}
{\hat {\cal H}}=-\sqrt{-G}\,\hat T^0_{~~0}= \frac{{\hat
\beta}_{p}{ {\cal T}}_{p}V(\hat T)}{ \sqrt{{\hat \beta}_{p}-{\hat
\alpha}_{p0}\dot{\hat T}^{2}}}.
\end{equation}
This first-order equation is consistent with the second-order NC tachyon
equation of motion, and, similar to Eq.~(\ref{eqn}), all possible
homogeneous NC rolling tachyons are given by one parameter
(${\hat {\cal H}}$) family of solutions of
Eq.~(\ref{nceq}). For any NC tachyon potential $V({\hat T})$
satisfying the runaway property $V({\hat T}=0)=1$ and $V({\hat
T}=\pm\infty)=0$, three types of NC rolling tachyons are obtained:
(i) When ${\cal {\hat H}}< \sqrt{\hat\beta_p}\,{ {\cal T}}_{p}$,
a convex-up (or down) NC
rolling tachyon connecting ${\hat T}(t=-\infty)=+\infty$ (or
$-\infty$) and ${\hat T}(t=\infty)=+\infty$ (or $-\infty$); (ii)
when ${\cal {\hat H}}=\sqrt{\hat\beta_p}\,{ {\cal T}}_{p}$,
 a monotonic increasing (or
decreasing) NC rolling tachyon connecting ${\hat T}(t=-\infty)=0$
and ${\hat T}(t=\infty)=+\infty$ (or $-\infty$) and; (iii) ${\cal
{\hat H}}>\sqrt{\hat\beta_p}\,{ {\cal T}}_{p}$,
 another monotonic increasing (or decreasing) NC
rolling tachyon connecting ${\hat T}(t=-\infty)=-\infty$ (or
$+\infty$) and ${\hat T}(t=\infty)=+\infty$ (or $-\infty$). Note
that (i) and (iii) are also named as full S-branes and (ii) is a
$\frac{1}{2}$S-brane~\cite{Gutperle:2002ai,Maloney:2003ck}. If we
choose the NC tachyon potential $V({\hat T})$ as $1/\cosh$ type
(\ref{V3}), the exact homogeneous NC rolling tachyon
solutions can be obtained as follows,
\begin{equation}\label{rs3}
\frac{{\hat \tau}(t)}{R}\equiv\sinh\left(\frac{{\hat T}(t)}{R}\right)= \left\{
\begin{array}{cl}
\sqrt{{\hat u}^{2}-1} \cosh\left(t/{\hat \zeta}\right) &
\mbox{for}~{\hat {\cal H}}<
\sqrt{\hat\beta_p}{ {\cal T}}_{p} \qquad ({\rm i})\\
\xi \exp \left(\pm t/{\hat \zeta}\right) &\mbox{for}~{\hat {\cal H}}=
\sqrt{\hat\beta_p}{ {\cal T}}_{p} \qquad ({\rm ii}) \\
\sqrt{1-{\hat u}^2} \sinh\left(t/{\hat \zeta}\right)
&\mbox{for}~{\hat {\cal H}}> \sqrt{\hat\beta_p}{ {\cal T}}_{p}
\qquad ({\rm iii}) \end{array}
\right. ,
\end{equation}
where ${\hat u}= \sqrt{\hat\beta_p}{ {\cal T}}_{p} / {\hat {\cal H}}$,
${\hat \zeta}=\sqrt{{\hat \alpha}_{p0}/{\hat \beta}_{p}}\, R$,
and $\xi$ is an arbitrary constant.

Though the functional forms of the exact NC rolling tachyon solutions (\ref{rs3})
correctly span all possible real solutions of linearized NC tachyon equation
in the background of constant NSNS two-form field
\begin{equation}
-\partial_{t}^{2}{\hat \tau}=-\frac{{\hat \beta}_{p}}{{\hat \alpha}_{p0}}
\frac{{\hat \tau}}{R^{2}},
\end{equation}
the degree of validity of an effective action can be judged by comparing
the obtained classical solutions (\ref{rs3}) of the EFT
with those of the open string theory, described by BCFT~\cite{Sen:2003bc}.
Specifically, this comparison is made with respect to the
energy density and pressure
given by matter responses to small gravitational fluctuations, and
current density of F1 given by responses to small NSNS two-form field
fluctuations:
\begin{equation}\label{src}
\delta \hat S = \frac{1}{2}\int d^{p+1}x\,\sqrt{-g}
\left( T^{\mu\nu} \delta g_{\mu\nu}+ J^{\mu\nu}\delta B_{\mu\nu}
\right).
\end{equation}

For clarity, let us take into account the single
electric component $E_0$ of the constant NSNS B-field along the $x$-direction
on a flat unstable D$p$-brane with closed string metric $\eta_{\mu\nu}$.
When the constant value of the energy density in BCFT is the same
one in the EFT
\begin{eqnarray}
T_{00} &=& \frac{{\cal T}_p \, V}{
\sqrt{G_0 - \dot{\hat T}^2}}\equiv T^{{\rm BCFT}}_{00}=
\left\{
\begin{array}{cl}
\frac{{\cal T}_p\cos^2(\pi\tilde \lambda)}{\sqrt{G_0}}\,
&  ({\rm i}) \\
\frac{{\cal T}_p}{\sqrt{G_0}} & ({\rm ii}) \\
\frac{{\cal T}_p\cosh^2(\pi\tilde\lambda)}{\sqrt{G_0}}\,
&  ({\rm iii})\end{array}\right., \label{eng2}
\end{eqnarray}
nonvanishing components of the pressure and the string charge density
for superstring theory are~\cite{Mukhopadhyay:2002en}
\begin{eqnarray}
T^{{\rm BCFT}}_{11} &=&\left\{ \begin{array}{cl}
-\frac{{\cal T}_p E_0^2
\cos^2(\pi\tilde\lambda)}{\sqrt{G_0}}
-\frac{{\cal T}_p \sqrt{G_0}
\cos^2(\pi\tilde\lambda)
\left[1 + \sin^2(\pi\tilde\lambda)\right]}{
\cos^4(\pi\tilde\lambda) + 4 \sin^2(\pi\tilde\lambda)
\cosh^2(\sqrt{G_0}t/\sqrt{2})}
& ({\rm i}) \\
-\frac{{\cal T}_p E_0^2}{\sqrt{G_0}}
-\frac{{\cal T}_p \sqrt{G_0}}{1 + 2\pi^2{\tilde\lambda}^2 e^{\sqrt{2G_0}t}}
& ({\rm ii}) \\
-\frac{{\cal T}_p E_0^2
\cosh^2(\pi\tilde\lambda)}{\sqrt{G_0}}
-\frac{{\cal T}_p \sqrt{G_0}\cosh^2(\pi\tilde\lambda)
\left[1-\sinh^2(\pi\tilde\lambda)\right]}{
\cosh^4(\pi\tilde\lambda) + 4 \sinh^2(\pi\tilde\lambda)
\sinh^2(\sqrt{G_0}t/\sqrt{2})}
& ({\rm iii})
\end{array}\right. , \label{prs3}\\
T^{{\rm BCFT}}_{ab} &=&
= \left(T^{{\rm BCFT}}_{11} + T^{{\rm BCFT}}_{00}E_0^2\right) \, \delta_{ab},
\qquad (a,b=2,3,\cdots, p), \label{prs4}\\
J_{{\rm BCFT}}^{01} &=& T^{{\rm BCFT}}_{00} E_0, \label{KR3}
\end{eqnarray}
where $\tilde{\lambda}$ is a parameter labelling the energy density
(\ref{eng2}) and
we have the following nonvanishing components of the open string metric
(\ref{omet}) and the NC parameter (\ref{ncpa})
\begin{eqnarray}
G_{0}&\equiv& 1-E_{0}^{2} =-G_{00}=G_{11}, \quad
G_{ab}=\delta_{ab},
\label{ome1}\\
\theta&\equiv& \frac{E_{0}}{1-E_{0}^{2}} =\theta^{01}=-\theta^{10},
\label{ncp1}\\
\hat \beta_p &=& G_0^2, \qquad \hat\alpha_{p0} = G_0.
\end{eqnarray}
Here, the cases (i), (ii), and (iii) correspond to the three cases in the
NC rolling tachyon solutions (\ref{rs3}).
By comparing Eq.~(\ref{nceq}) with Eq.~(\ref{eng2}),
following relation for the pure electric case with $E_{0}$ along
$x$-direction is
\begin{eqnarray}
\hat {\cal H} =
\left\{ \begin{array}{cl}
{\cal T}_p G_{0}\cos^{2}(\pi\tilde\lambda) & ({\rm i}) \\
{\cal T}_p G_{0} & ({\rm ii}) \\
{\cal T}_p G_{0}\cosh^{2}(\pi\tilde\lambda) & ({\rm iii})
\end{array}\label{val}
\right. ,
\end{eqnarray}
and the nonvanishing pressure components and string charge density are
\begin{eqnarray}
 T_{11} &=& -T_{00}(1- \dot{\hat T}^2) \nonumber \\
&=&\left\{ \begin{array}{cl} -\frac{{\cal T}_p E_0^2
\cos^2(\pi\tilde\lambda)}{\sqrt{G_0}}
- \frac{{\cal T}_p \sqrt{G_0} \cos^2(\pi\tilde\lambda)}{
\cos^4(\pi\tilde\lambda) + \sin^2(\pi\tilde\lambda) \left[
1 + \cos^2(\pi\tilde\lambda)\right] \cosh^2\left(\sqrt{G_0}\,
t/R\right)} &  ({\rm i}) \\
-\frac{{\cal T}_p E_0^2}{\sqrt{G_0}}- \frac{{\cal T}_p
\sqrt{G_0}}{1 + \xi^2 e^{2\sqrt{G_0}\, t/R}} &
 ({\rm ii}) \\
-\frac{{\cal T}_p E_0^2
\cosh^2(\pi\tilde\lambda)}{\sqrt{G_0}}
- \frac{{\cal T}_p \sqrt{G_0} \cos^2(\pi\tilde\lambda)}{
\cosh^4(\pi\tilde\lambda) + \sinh^2(\pi\tilde\lambda) \left[
1 + \cosh^2(\pi\tilde\lambda)\right] \sinh^2\left(\sqrt{G_0}\,
t/R \right)} &  ({\rm iii})
\end{array}\right. ,\label{prs1}\\
T_{ab} &=& -T_{00}(G_0- \dot{\hat T}^2) =
\left(T_{11} + T_{00}E_0^2\right) \, \delta_{ab},
\label{prs2}\\
J^{01} &=& T_{00} E_0.
\label{KR2}
\end{eqnarray}
If there is no contribution from the F1 represented by nonvanishing
electric field $E_{0}$, the NC tachyon field has $\lim_{t\rightarrow \infty}
\dot{\hat T}^2 \rightarrow 1$ and thereby, the homogeneous tachyon matter
becomes pressureless, $\lim_{t\rightarrow \infty} T_{11}
=\lim_{t\rightarrow \infty} T_{ab} \rightarrow 0$, as long as the production of
closed string degrees is neglected.

Due to the equivalence (\ref{equ}) between the effective action (\ref{fa})
and the NC action (\ref{nct}), the pressure (\ref{prs1})--(\ref{prs2})
and the F1 charge density (\ref{KR2}) coincide exactly with those of
EFT as a defining property~\cite{Kim:2003he}. In addition,
the F1 charge density (\ref{KR2}) is the same as that in BCFT (\ref{KR3}).
The relative coefficients of $T_{11}$ and $T_{ab}$ computed from
EFT or NCFT in Eqs.~(\ref{prs1}) and (\ref{prs2}) agree qualitatively to
but do not coincide exactly with
those of BCFT in Eqs.~(\ref{prs3}) and (\ref{prs4}) for the full S-brane
cases (i) and (iii).
This mismatch disappears for the half S-brane case (ii); that is,
if we choose the origin of time as that satisfying
$\xi = \sqrt{2}\pi\tilde\lambda$,
the stress tensors share the same results with BCFT.
This coincidence is also satisfied for the half S-brane in bosonic string theory
with identification $\xi= \sqrt{2\pi\tilde\lambda}\,$.
The coincidence looks natural since
the effective action (\ref{fa}) and the tachyon potential (\ref{V3})
of our consideration have been derived by open string theory
by taking into account
the fluctuations around half S-brane configuration
with the higher derivatives neglected, i.e., $\partial^2 T = \partial^3 T=
\cdots = 0$~\cite{Kutasov:2003er}.
Though the pure electric case is compared,
its generalization to arbitrary constant NSNS two-form field is
automatic~\cite{Rey:2003xs}.
Comparing the obtained results to those from the string field theory~\cite{Fujita:2004ha}
or $c=1$ matrix model~\cite{McGreevy:2003kb} is not yet made at the level of
energy-momentum tensor.

\setcounter{equation}{0}
\section{NC Rolling Tachyon with NC Electric Field}
U(1) gauge field lives on a D-brane and, in the context of NCFT,
it appears as an NC U(1) gauge field
${\hat F}_{\mu\nu}$ on our single unstable D-brane.
Without tachyonic degree the equivalence between the DBI-type action
in ordinary field theory and the NCFT (\ref{SW}) still holds
with inclusion of the NC U(1) gauge field.
Various tachyon actions
proposed~\cite{Dasgupta:2000ft,Banerjee:2004cw,Kim:2004xn}, and, for those
actions,
reduction to the action (\ref{nct}) in vanishing ${\hat F}_{\mu\nu}$ limit
is a necessary condition.
For the following DBI-type action~\cite{Banerjee:2004cw,Kim:2004xn},
EFT and NCFT including the slowly-varying
NC U(1) gauge field and NC tachyon field were proven to be equivalent
up to the leading order of
NC parameter $\theta^{\mu\nu}$
\begin{equation}\label{nctf}
{\hat S}_{{\hat F}}=-\frac{{\hat {\cal T}}_p}{2} \int d^{p+1}x\;
\left[\hat{V}(\hat{T}) \ast \sqrt{ -\det {}_{\ast}(G_{\mu\nu}+{\hat F}_{\mu\nu}
+{\hat D}_{\mu}\hat{T}\ast{\hat D}_{\nu}\hat{T})}
+(\sqrt{\hspace{-5mm}}\leftrightarrow \hat{V}) \right],
\end{equation}
where
\begin{equation}
{\hat D}_{\mu}\hat{T}=\partial_{\mu}{\hat T}
-i({\hat A}_{\mu}\ast {\hat T}-{\hat T}\ast {\hat A}_{\mu}).
\end{equation}

Here, let us restrict our interest to pure electric cases.
For a flat unstable D-brane of arbitrary $p$ with a constant
electric component of NSNS two form field $B_{0i}={\bf E}_{0}^{i}$,
direction of the electric component can be always chosen as the
$x$-direction. Then we have closed string variables
$\eta_{\mu\nu}$ and ${\bf E}_{0}=
E_{0}{\bf x}$, and the corresponding open string
variables, $G_{\mu\nu}$ and $\theta^{\mu\nu}$, are given in
Eqs.~(\ref{ome1})--(\ref{ncp1}).

In the homogeneous NC rolling tachyon configurations,
the NC tachyon and NC U(1) gauge field on the brane depend only on time
${\hat T}={\hat T}(t)$ and ${\hat F}_{\mu\nu}={\hat F}_{\mu\nu}(t)$.
As we assumed, the field strength tensor has
only electric components, ${\hat F}_{0i}={\hat {\bf E}}^{i}(t)$ and
${\hat F}_{ij}=0$. At a given time $t$, two vectors ${\bf E}_{0}$
(or equivalently $\vec{\theta}^{1}=\theta {\bf x}$) and ${\hat {\bf E}}$
generally form a plane, and we work in the coordinates where the
NSNS field is directed in the $x$-direction and the NC electric field
lies on the $xy$-plane
\begin{equation}\label{ean}
{\hat F}_{01}={\hat E}_{1}(t),\quad {\hat F}_{02}={\hat E}_{2}(t).
\end{equation}

In order to proceed, we choose a gauge
\begin{equation}\label{gau}
\theta^{0\mu}{\hat A}_{\mu}=0
\end{equation}
which leads to ${\hat A}_{1}=0$ since
$\theta=\theta^{01}$ is nonvanishing among all the NC parameters.
Substituting the gauge (\ref{gau}) into the definition of NC field
strength tensor (\ref{ncF}),
\begin{equation}\label{e12}
{\hat E}_{1}=\partial_{1}{\hat A}_{0},\qquad
{\hat E}_{2}=(1+{\hat E}_{1}\theta)\partial_{0}{\hat A}_{2}
-\partial_{2}{\hat A}_{0}.
\end{equation}
Note that the NC electric field (\ref{e12}) automatically satisfies NC Bianchi
identity ${\hat D}_{\mu}{\hat F}_{\nu\rho}+
{\hat D}_{\nu}{\hat F}_{\rho\mu}+{\hat D}_{\rho}{\hat F}_{\mu\nu}
=0$~\cite{Banerjee:2003vc,Kim:2004xn}.

If we summarize our results under our ansatz (\ref{ide}) and (\ref{ean}),
every star product is replaced by an ordinary product and then
${\hat D}_{\mu}{\hat T}=\delta_{\mu 0}(1+{\hat E}_{1}\theta)
\dot{{\hat T}}^{2}$. Inserting the results into the action (\ref{nctf}),
we obtain a simplified form of it
\begin{equation}\label{eac}
{\hat S}_{{\hat F}}=-{\hat {\cal T}}_{p}\int dt dx dy d^{p-2}x_{\perp}
{\hat V}({\hat T})\sqrt{-{\hat X}}\, ,
\end{equation}
where $x_{\perp}$s denote transverse coordinates and
\begin{equation}\label{xx}
-{\hat X}\equiv \beta_{{\hat F}}-\alpha_{0{\hat F}}\dot{{\hat T}}^{2}
=G_{0}^{2}-G_{0}{\hat E}_{2}^{2}-{\hat E}_{1}^{2}
-G_{0}(1+{\hat E}_{1}\theta)^{2}\dot{{\hat T}}^{2}.
\end{equation}
From the action (\ref{eac}), the survived components of the equations
of motion are the time-component of the conservation of energy-momentum tensor
\begin{equation}\label{em0}
(1+{\hat E}_{1}\theta)\partial_{0}\left[
\frac{{\hat {\cal T}}_{p}{\hat V}}{\sqrt{-{\hat X}}} \right] =0,
\end{equation}
and $x$- and $y$-components of gauge field equations are
\begin{equation}\label{ga12}
(1+{\hat E}_{1}\theta)\partial_{0}\left[
\frac{{\hat {\cal T}}_{p}{\hat V}}{\sqrt{-{\hat X}}}{\hat E}_{1}\right]=0,
\qquad
(1+{\hat E}_{1}\theta)\partial_{0}
\left[\frac{{\hat {\cal T}}_{p}{\hat V}}{\sqrt{-{\hat X}}}G_{0}{\hat E}_{2}\right]=0.
\end{equation}
Solving the equations (\ref{em0})--(\ref{ga12}) when $1+{\hat E}_{1}\theta
\ne 0$, we show that the NC electric field ${\hat {\bf E}}$ is constant
and the only nontrivial
equation for the NC tachyon reduces to
\begin{equation}\label{gam}
{\hat \gamma}=\frac{{\hat {\cal T}}_{p}{\hat V}}{\sqrt{-{\hat X}}}
={\rm constant}.
\end{equation}
The resultant equation (\ref{gam}) is formally the same as Eq.~(\ref{nceq})
under identification, ${\hat \gamma}\Leftrightarrow {\hat {\cal H}}$,
$\beta_{{\hat F}}\Leftrightarrow {\hat \beta}_{p}$, ${\hat \alpha}_{0{\hat F}}
\Leftrightarrow \alpha_{p0}$, and ${\hat T}\Leftrightarrow {\hat T}$,
so that the spectrum of homogeneous NC rolling tachyons is exactly the same
as that in the previous section as given between Eq.~(\ref{nceq}) and
Eq.~(\ref{rs3}). Here let us omit displaying the solutions in Eq.~(\ref{rs3})
again.
For the NC rolling tachyon solutions, there are constant nonvanishing
electric fluxes
\begin{eqnarray}
\hat \Pi^1 = \hat\gamma \hat E_1, \qquad
\hat \Pi^2 = \hat\gamma G_0\hat E_2,
\end{eqnarray}
which are defined by ${\hat \Pi}^{i}=\delta {\hat S}/\delta {\hat E}_{i}$
and signify the existence of F1 string fluid. The corresponding
energy-momentum tensor has constant energy density ${\hat T}^{00}=\hat \gamma$,
nonvanishing constant stress along $12$-direction ${\hat T}^{12} =
-\frac{\hat\gamma}{G_0}\hat E_1\hat E_2$, and zero momentum density
${\hat T}^{0i} =0$. On the other hand, all the pressure components are
nonvanishing in this reference frame
\begin{eqnarray}
\hat T^{11} &=& \frac{\hat\gamma}{G_0}
\left[-G_0 + \hat E_2^2 + (1 + \hat E_1\theta)^2 \dot{{\hat T}}^2\right],\\
\hat T^{22} &=& \frac{\hat\gamma}{G_0}
\left[-G_0^2 +\hat E_1^2 + G_0(1 + \hat E_1\theta)^2 \dot{{\hat T}}^2\right],\\
\hat T^{ij} &=& \frac{\hat\gamma}{G_0} {\hat X}\, \delta^{ij},
\qquad (i,j = 3,4,\cdots, p).
\end{eqnarray}

The functional form of the NC rolling tachyon solutions is
the same as that of rolling tachyons in EFT, but their equivalence in the action
level is guaranteed only up to the leading NC parameter. Therefore, the comparison of
the components of energy-momentum tensor given by responses to small
gravitational fluctuations $\delta g_{\mu\nu}$ in NCFT to those in EFT and
BCFT can be valid, at most, up to the leading NC parameter.
In the analysis of the NC rolling tachyon by the use of different NC tachyon
actions~\cite{Dasgupta:2000ft}, all three species of the NC rolling tachyon
cannot be obtained as were obtained in the case of NC tachyon kinks~\cite{Kim:2004xn}.

\setcounter{equation}{0}
\section{Summary}
We considered a DBI type NCFT action of a real NC tachyon
field ${\hat T}$ as an EFT describing dynamics of a flat unstable D$p$-brane.
When the background of constant NSNS two-form field is coupled to the brane,
this field is expressed in terms of a constant open string metric and NC parameter.
For the general constant open string metric and NC parameter,
we showed the existence of all possible homogeneous NC rolling tachyon solutions
irrespective of specific shape of the NC tachyon potential ${\hat V}$,
once ${\hat V}({\hat T}=0)=1$ and ${\hat V}({\hat T}=\pm\infty)=0$
are satisfied.
These solutions are classified into two full S-brane solutions, one with
${\hat T}(t=-\infty)=\pm\infty$ and ${\hat T}(t=+\infty)=\mp\infty$
and the other with ${\hat T}(t=-\infty)=\pm\infty$
and ${\hat T}(t=+\infty)=\pm\infty$, and one half S-brane with
${\hat T}(t=-\infty)=0$ and ${\hat T}(t=+\infty)=\pm\infty$.
Under a specific $1/\cosh$ type NC tachyon potential, they are obtained
as exact solutions whose functional forms are hyperbolic sine,
hyperbolic cosine, and exponential, respectively, which are
classical configurations of a linearized tachyon equation in BCFT.
When the energy-momentum tensor and the F1 charge density in NCFT are compared with
those in BCFT, pressure and stress components agreed qualitatively
but not exactly to those in BCFT for the full S-branes but coincided
exactly to those for the half S-brane. This result seems natural since the $1/\cosh$ type
potential in EFT was derived by considering the fluctuations around
half S-brane in open string theory.

The addition of DBI type electromagnetism preserves the structure of
NCFT including the covariant derivative of the NC tachyon. However, the
allowed configuration of field strengths was constant,
and the obtained homogeneous NC rolling tachyons had the same
functional form.

\section*{Acknowledgements}
The authors would like to thank Chong Oh Lee for valuable discussions.
This work is the
result of research activities (Astrophysical Research Center for
the Structure and Evolution of the Cosmos (ARCSEC)) supported by
the Korea Science $\&$ Engineering Foundation.

\end{document}